\documentclass[iop,ppcf,superscriptaddress,reprint,nofootinbib]{revtex4-1}

\usepackage{amsfonts}
\usepackage{mathrsfs}  
\usepackage[pdftex]{graphicx}
\usepackage{graphicx}
\usepackage{bm}%
\usepackage{amsmath} 
\usepackage[caption=false]{subfig}
\usepackage{color}
\usepackage{hyperref}
\usepackage{soul}
\usepackage{ulem}
\usepackage[squaren]{SIunits}

\usepackage{booktabs}

\let\vec\boldsymbol

\begin{document}
\title{Electron spin polarization in realistic trajectories around the magnetic node of two counter-propagating, circularly polarized, ultra-intense lasers}
\author{D. Del Sorbo}
\affiliation{York Plasma Institute, Department of Physics, University of York, York YO10 5DD, United Kingdom}
\author{D. Seipt}
\affiliation{Cockcroft Institute, Daresbury Laboratory, Warrington WA4 4AD, United Kingdom}
\affiliation{Department of Physics, Lancaster University, Lancaster LA1 4YB, United Kingdom}
\author{A. G. R. Thomas}
\affiliation{Department of Physics, Lancaster University, Lancaster LA1 4YB, United Kingdom}
\affiliation{Center for Ultrafast Optical Science, University of Michigan, Ann Arbor, Michigan 48109-2099, USA}
\author{C. P. Ridgers}
\affiliation{York Plasma Institute, Department of Physics, University of York, York YO10 5DD, United Kingdom}

\begin{abstract}
It has recently been suggested that two counter-propagating, circularly polarized, ultra-intense lasers can induce a strong electron spin polarization at the magnetic node of the electromagnetic field that they setup \cite{del2017spin}. We confirm these results by considering a more sophisticated description that integrates over realistic trajectories. The electron dynamics is weakly affected by the variation of power radiated due to the spin polarization. The degree of spin polarization differs by approximately 5\% if considering electrons initially at rest or already in a circular orbit. The instability of trajectories at the magnetic node induces a spin precession associated with the electron migration that establishes an upper temporal limit to the polarization of the electron population  of about one laser period. 
\end{abstract}

    \maketitle

    \section{Introduction}
    
  Since the application of chirped pulse amplification to optical laser pulses \cite{strickland1985compression}, laser intensities have increased dramatically, surpassing $10^{22}$ W/cm$^{2}$ \cite{yanovsky2008ultra}. The increase in the electromagnetic fields in the laser focus, resulting from the increase in laser intensity, has enabled the investigation of new regimes in laser-produced plasmas.  For example, as laser intensities increased beyond $10^{19}$   W/cm$^{2}$, plasma electrons became relativistic \cite{gibbon2004short,umstadter1996nonlinear}, paving the way to new applications, such as laser driven particle acceleration \cite{macchi2013ion,esarey2009physics,ghotra2016multi,ghotra2016polarization}.\color{black}

Several facilities being constructed as part of the Extreme Light Infrastructure project (ELI) \cite{korn2013eli}  aim to surpass a new laser intensity threshold ($I\gtrsim5\times10^{23}$  W/cm$^{2}$).  At this intensity, strong-field quantum-electrodynamics (QED) effects \cite{andersen2012experimental,turcu2016high} are expected to play an important role in the collective plasma dynamics \cite{bell2008possibility,ridgers2012dense,zhang2015effect}. This new regime -- so called QED plasma -- is inferred to exist also in extreme astrophysical environments, such as the pulsar magnetosphere \cite{goldreich1969pulsar} and the black hole dyadosphere \cite{ruffini2010electron}.

The important QED effects, expected to play a major role in laser-created QED plasmas, are \cite{RevModPhys.38.626,di2012extremely,Ritus:JSLR1985,seipt2016depletion}: (i) incoherent emission of MeV energy gamma-ray photons by electrons and positrons on acceleration by the macroscopic electromagnetic fields in the plasma (strongly non-linear Compton scattering), with the resulting radiation-reaction (RR) strongly modifying the dynamics of the emitting electron or positron \cite{ridgers2014modelling,di2010quantum}; (ii) pair creation by the emitted gamma-ray photons, in the macroscopic electromagnetic fields (the multi-photon Breit-Wheeler process \cite{burke1997positron}). Moreover, processes involving other particles such as muons and pions,
but also more exotic particles like axions may appear as well \cite{di2012extremely,CAST:NaturePhys2017}.

The first steps toward experimental tests of the existing theory \cite{baier1968processes,ridgers2014modelling, ridgers2017signatures}  of quantum RR have been recently performed  \cite{poder2017evidence,cole2017experimental, wistisen2017experimental} but the role of the fermion spin has received relatively little investigation \cite{PhysRevE.96.023207,del2017spin,marklund2007dynamics,brodin2007spin2,brodin2007spin,brodin2010spin,ahrens2017electron}. While radiating gamma-rays via non-linear Compton scattering, electrons may undergo spin-flip transitions. It is well known that for an electron orbiting in a constant magnetic field, spin-flip transitions where the final projection of the spin onto the axis defined by the magnetic field is antiparallel to the magnetic field are more energetically favorable than the reverse \cite{sokolov1966synchrotron}. Furthermore, it has been recently shown that the same is true for an electron orbiting with normalized velocity $\boldsymbol{\beta}$ in a rotating electric field $\boldsymbol{E}$, where the vector $\boldsymbol{E}\times\boldsymbol{\beta}$ plays the same role as the magnetic field \cite{del2017spin}.  The latter case is potentially experimentally realizable at the magnetic node of the standing wave formed by two counter-propagating circularly-polarized laser pulses.  By considering highly idealized stationary orbits, it has been shown  that electrons at the magnetic node should rapidly spin polarize. Similar spin flip transitions can occur when energetic electrons radiate in the strong atomic fields as they pass down the axis of a crystal lattice.  It has been shown that spin flip transitions, at the expected rate, are required to reproduce the emitted gamma-ray spectrum measured experimentally \cite{kirsebom2001first}.

In this article we confirm the predictions of Ref.~\cite{del2017spin}, considering a more detailed description of time-dependent trajectories, obtained by the numerical integration of the electron equations of motion, coupled to the equations that describe the spin dynamics. The influence of different parameters such as the RR force, the initial phase space configuration and the effect of electron migration away from the magnetic node, which is an unconditionally unstable point, are analyzed.

The article is organized as follows. In Sec.~\ref{Deterministic electron spin dynamics}, the dynamics of spin polarized electrons is described, detailing the equations that have to be numerically solved in order to track the electron trajectory, the spin polarization direction and the probability of spin polarization. In Sec.~\ref{Electron trajectory spin polarization}, the influence of electron trajectory on the degree of spin polarization is investigated, considering the effects of the RR force and the effects of the instability of trajectories at the magnetic node. In Sec.~\ref{Discussion}, the derived results and their implications are discussed. Finally, in Sec.~\ref{Conclusions} conclusions are drawn.

    \section{Spin polarized electron dynamics}\label{Deterministic electron spin dynamics}
    
 Classically, a free electron in an electromagnetic field is subjected to acceleration by the Lorentz force. On acceleration, the electron emits electromagnetic radiation and, consequently, loses energy. The reaction of the electron to the radiation it emits can be modeled as an additional RR force $\boldsymbol{F}_{RR}$. Therefore, the electron's equation of motion is 
      \begin{equation}
   \frac{d\boldsymbol{p}}{dt}=-e(\boldsymbol{E}+\boldsymbol{\beta}\times \boldsymbol{B})+\boldsymbol{F}_{RR},\label{lorentz frce}
   \end{equation}
   with $\boldsymbol{p}=\gamma m_{e} c \boldsymbol{\beta}$ ($\gamma$ is the Lorentz factor and $\boldsymbol{\beta}=\boldsymbol{v}/c$). The constants $m_{e}$, $e$ and  $c$ are the electron mass, the elementary charge and the speed of light while the vectors $\boldsymbol{E}$ and $\boldsymbol{B}$ represent the macroscopic electric and magnetic fields. A simple form for the classical RR force can be derived from the Lorentz-Abraham-Dirac equation by using the Landau-Lifshitz approach 
   \cite{landau2013classical,Klepikov:PhysUspekh1985,tamburini2010radiation,Niel2017}.
In the ultra-relativistic limit it reads   
   \cite{landau2013classical,ridgers2017signatures}
   \begin{equation}
   \boldsymbol{F}_{RR}=-\frac{\mathcal{P}_{c}}{c}\frac{\boldsymbol{p}}{\left\lVert\boldsymbol{p}\right\rVert},\label{classical_RR}
   \end{equation}
   where
    \begin{equation}
    \mathcal{P}_{c}=\frac{4\pi m_{e}c^{3}}{3\lambda_{C}}\alpha_{f} \eta^{2}
    \end{equation}
    is the power radiated classically.
    $\lambda_{C}={2\pi \hbar}/({m_{e}c})$ is the Compton wavelength, 
    $\hbar$ is Planck's constant, 
     and $\alpha_{f} =e^{2}/(\hbar c)$ is the fine-structure constant.
    The quantum parameter
    \begin{equation}
    \eta=\frac{e\hbar\gamma}{m_{e}^{2}c^{3}}\sqrt{\left(\boldsymbol{E}+\boldsymbol{\beta}\times\boldsymbol{B}\right)^{2}-\left(\boldsymbol{\beta}\cdot\boldsymbol{E}\right)^{2}}
    \end{equation}
    measures the magnitude of quantum effects such as electron recoil due to photon emission in non-linear Compton scattering \cite{di2012extremely}. 
    
	Equation \eqref{classical_RR} neglects the quantum nature of RR force, according to which,
	the electron radiation is the sum of successive incoherent and stochastic gamma-ray emissions.
	We can account for part of the quantum effects (the reduction in the radiated power due to electron recoil, although not stochasticity)
	 \cite{sokolov1966synchrotron,kirk2009pair}
	by multiplying $ \mathcal{P}_{c}$ by the Gaunt factor $g(\eta)$, that represents the
	ratio of the quantum to classically radiated power. The RR force then becomes
   \begin{equation}
   \boldsymbol{F}_{RR}=-\frac{g(\eta)\mathcal{P}_{c}}{c}\frac{\boldsymbol{p}}{\left\lVert\boldsymbol{p}\right\rVert}.\label{QRR}
   \end{equation}
A convenient fit to $g$ is \cite{baier1991quasiclassical}
	\begin{equation}
        \label{old_gaunt}
	g(\eta)\approx\left[1 + 4.8\left(1 +\eta\right) \ln\left(1 + 1.7\eta\right) + 2.44\eta^{2}\right]^{-2/3}.
	\end{equation}
	This form of RR gives a good approximation to the average energy loss by an ensemble of electrons \cite{ridgers2017signatures,niel2017quantum}.
	
Another quantum aspect of radiation is that during emission of a gamma-ray photon the electron's spin may flip.  To describe the evolution of the electron's spin, we consider the spin expectation value vector $\boldsymbol{S}$.

The classical evolution of $\boldsymbol{S}$ can be described by the Bargmann-Michel-Telegdi (BMT) equation \cite{bargmann1959precession,vieira2011polarized}
\begin{equation}
\begin{split}
\frac{d\boldsymbol{S}}{d\tau}=& -\frac{g_{e}}{2} \frac{e}{m_{e}c}
 \left(\boldsymbol{S}\times\boldsymbol{B}  +  \boldsymbol{E}\boldsymbol{S}\cdot\boldsymbol{\beta}\right)
\\
& -
\frac{g_{e}}{2}\frac{e\gamma^{2}\boldsymbol{\beta}}{m_{e}c}
\left(\boldsymbol{S}\cdot\boldsymbol{\beta}\boldsymbol{E}\cdot\boldsymbol{\beta}   -  \boldsymbol{E}\cdot\boldsymbol{S}  +\boldsymbol{S}\times\boldsymbol{B}\cdot\boldsymbol{\beta}\right)
\\
&-\frac{\gamma^{2}\boldsymbol{\beta}}{m_{e}c}
\left(\boldsymbol{S}\cdot\boldsymbol{\beta}\frac{d\boldsymbol{p}}{dt}\cdot\boldsymbol{\beta}   -  \frac{d\boldsymbol{p}}{dt}\cdot\boldsymbol{S}  \right)
,\label{BMT eq}
\end{split}
\end{equation}
where $\tau=t/\gamma$ is the proper time and 
$(g_{e}-2)/2\approx 1.16\times10^{-3}$ 
is the electron anomalous magnetic moment\footnote{The electron anomalous magnetic moment may be modified by strong field interactions \cite{meuren2011quantum,ritus1979quantum}. However, this is a second order correction for $\eta\lesssim1$ and, even for $\eta\gtrsim1$ should not affect the qualitative description of spin precession. }.
Equation \eqref{BMT eq}
 gives a general version of the BMT equation proposed in Eq.~(6) of Ref.~\cite{bargmann1959precession}, which
is independent of the way in which one defines the force (${d\boldsymbol{p}}/{dt}$) acting on the  electron.
 This is useful as $d\boldsymbol{p}/dt$ depends on the model used to describe the RR force,
see also Refs.~\cite{Kar:AnnPhys2011,Wen:PRA2017}.
Another aspect of electron spin is the fact that Stern-Gerlach forces can affect the electron's motion, however this
is a factor of $\omega/(\alpha_f m_{e}\eta)$ smaller than the RR force \cite{tamburini2010radiation} and is therefore neglected.

The BMT equation describes the classical precession of the electron's spin between emission events.  If the spin-basis does not precess in time, i.e. $d\boldsymbol{\zeta}/d\tau=0$, then polarization along $\boldsymbol{\zeta}$ is preserved over the classical trajectory between emission events.  The polarization in this direction may then only change by spin-flip transitions during emission. \color{black}

 In the electron's instantaneous rest frame, $\boldsymbol{\zeta}$ is always parallel to the magnetic field calculated in this frame.
The probability $P^{s}$ of the electrons being spin polarized parallel ($s=\uparrow=+1$) or antiparallel  ($s=\downarrow=-1$) to $\boldsymbol{\zeta}$ obeys to the following master equations \cite{del2017spin}:
\begin{equation}
\begin{split}
\frac{dP^{\uparrow}}{d\tau}=P^{\downarrow}\frac{dN^{\downarrow\uparrow}}{d\tau}-P^{\uparrow}\frac{dN^{\uparrow\downarrow}}{d\tau},\\
\frac{dP^{\downarrow}}{d\tau}=P^{\uparrow}\frac{dN^{\uparrow\downarrow}}{d\tau}-P^{\downarrow}\frac{dN^{\downarrow\uparrow}}{d\tau},
\end{split}\label{poability evolution}
\end{equation}
where ${dN^{ss'}}/{d\tau}$ is the rate of gamma-ray emission with spin flip transition from $s$ to $s'$, whose  explicit form is provided in Eq.~(3) of Ref.\cite{del2017spin}.

It has been shown \cite{del2017spin} that spin polarization of the electron population can modify the power radiated by up to 20\% (as some transitions available to an unpolarized population of electrons are no longer possible) and, consequently, the Gaunt factor.
Here we fit the spin dependent Gaunt factor $g^s(\eta)$ for $s=\pm1$ as
\begin{multline} \label{gaunt-spin}
g^s(\eta) \approx 
\left[ 1 +   2.54 ( \eta^2 - 1.28 s \eta   ) \right. \\
\left.
        + ( 4.34 + 2.58 s  )  (1 +  \eta) \ln( 1 + (1.98 +  0.11 s)\eta )           
 \right]^{-2/3} \,,
\end{multline}
which in the case where $s=0$ gives a more precise fit of the spin-independent Gaunt factor to that given in equation (\ref{old_gaunt}).
 We may therefore include the effect of spin in the RR force as 
   \begin{equation}
   \boldsymbol{F}_{RR}=-\frac{\left[P^{\downarrow}g^{\downarrow}(\eta)+P^{\uparrow}g^{\uparrow}(\eta)\right]\mathcal{P}_{c}}{c}\frac{\boldsymbol{p}}{\left\lVert\boldsymbol{p}\right\rVert}.\label{SRR}
   \end{equation}

 This description of spin dynamics relies on the existence of a globally non-precessing spin basis and, therefore, is limited to particular field configurations. In more general electromagnetic field configurations it is usually not possible to identify a globally non-precessing spin-basis and there is as yet no way to describe the spin dynamics during multiple photon emissions.
\color{black}

In this article, we focus on a particular laser configuration \cite{bell2008possibility}: counter-propagating, circularly polarized plane-waves which produce a standing wave of the form
\begin{equation}
\begin{split}
\boldsymbol{E}=\frac{a_{0}m_{e}c\omega}{e}
\left( \begin{array}{c}
\cos(kz)\cos(\omega t) \\
\cos(kz)\sin(\omega t) \\
0 \end{array} \right),
\\
\boldsymbol{B}=\frac{a_{0}m_{e}c\omega }{e}
\left( \begin{array}{c}
\sin(k z)\cos(\omega t) \\
\sin( k z)\sin(\omega t) \\
0 \end{array} \right),
\end{split}\label{laser field}
\end{equation}
    where the electromagnetic waves propagate parallel and antiparallel to the $z$-direction, $\omega =2\pi c/\lambda $ is the laser frequency, $k =\omega /c$ is the magnitude of the laser wavevector and $a_{0}\approx 85.5\lambda /\mu\mbox{m}\sqrt{I/(10^{22}\;\mbox{W/cm}^{2})}$ is the laser's strength parameter, (where $I$ and $\lambda$ are the intensity and wavelength of each laser and equal in this case). Typically, for high intensity lasers, $\lambda  \approx 1\;\mu$m and the laser period is $T=\lambda/c\approx3.33$ fs.
    
\begin{figure}
        \centering
        \includegraphics[width=0.6\columnwidth]{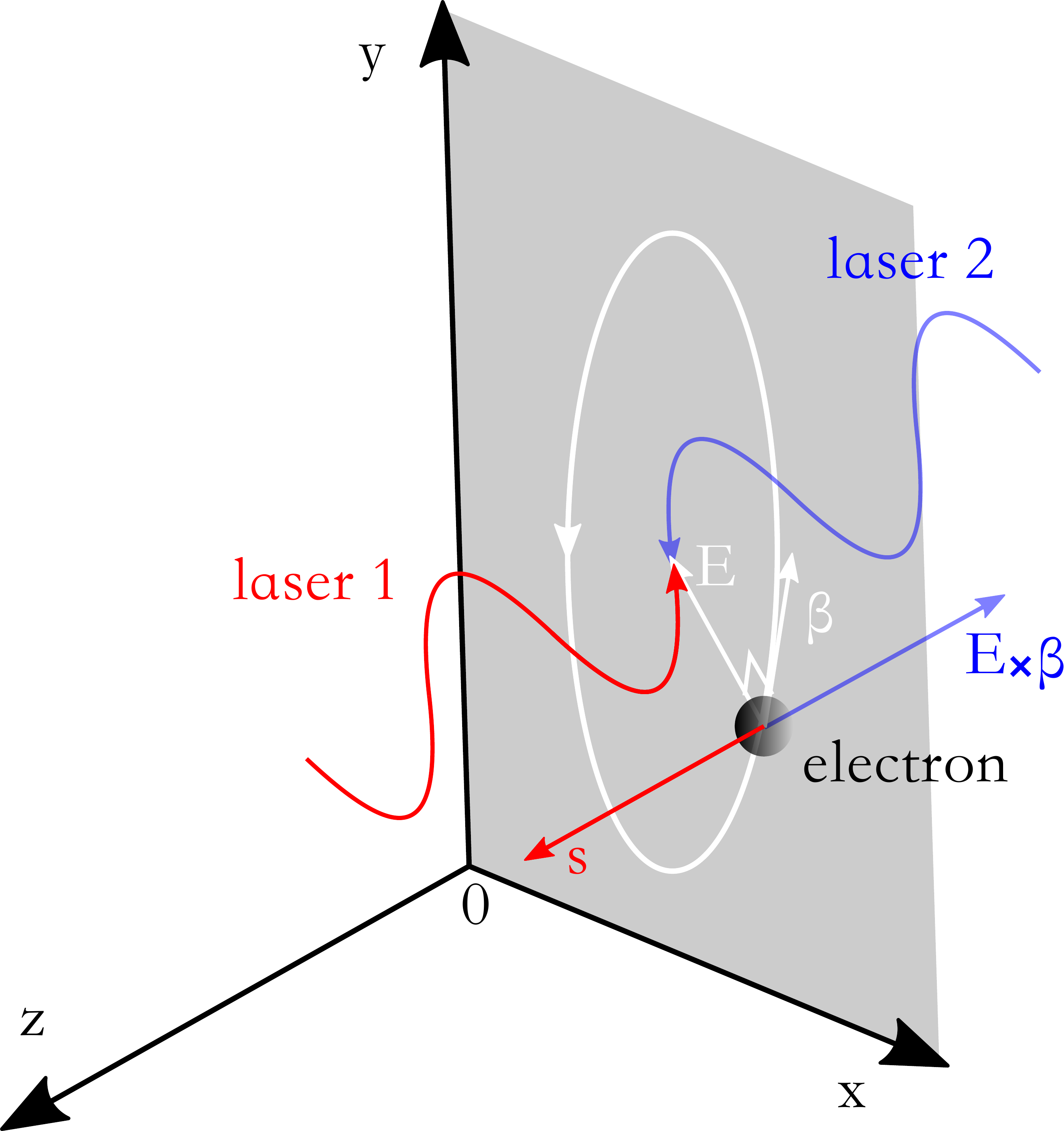}
      \caption{
       Schematic representation of the electron spin polarization described in this article. A standing wave is produced by the counter-propagation of two circularly polarized lasers. At the magnetic node ($z=0$) the electric field
        $\boldsymbol{E}$ 
        rotates with a constant amplitude, inducing the rotation of any electron in this plane. 
        In such a trajectory, the electron tend to align its spin $s$ antiparallel to the vector $\boldsymbol{E} \times \boldsymbol{\beta} $. \label{schematic}}
\end{figure}

    In the plane of the magnetic nodes,
                 i.e.~where $kz$ is an integer multiple of $2\pi$,
     the electromagnetic fields given in Eq.~\eqref{laser field}
    reduce to a rotating electric field. In this particular plane, the spin polarization direction (non-precessing spin basis) is $\boldsymbol{\zeta}=\boldsymbol{E}\times\boldsymbol{\beta}/\left\lVert\boldsymbol{E}\times\boldsymbol{\beta}\right\rVert=\boldsymbol{e}_{z}$, where $\boldsymbol{e}_{z}=(0,0,1)$    \footnote{In Ref.~\cite{raicher2016nonlinear}, a correction for gamma-ray emission rates in the rotating electric field configuration was proposed. However, the latter is derived for scalar fields (spin-0 electrons) and does not apply to ultra-relativistic particles, as considered in this article.}. 
We are going to show that electrons tend to spin polarize antiparallel to $\boldsymbol{\zeta}$, as shown schematically in Fig.~\ref{schematic}.
Outside the magnetic node plane, it is not possible to identify a non-precessing spin basis. Therefore, beyond the magnetic node plane it is not possible to perform any prediction of electron spin polarization using the current model. 


The standing wave set up by two counter-propagating circularly polarized lasers is a configuration favorable for the observation of strong-field QED effects \cite{bell2008possibility,grismayer2017seeded,kirk2009pair,kirk2016radiative}. Moreover, it is a simple case because there exists a globally non-precessing spin basis.
\color{black}

\section{Realistic trajectories \& spin polarization}\label{Electron trajectory spin polarization}

We have numerically solved Eq.~\eqref{lorentz frce} in order to track the electron trajectory in the particular laser configuration considered (the magnetic node of two circularly polarized counter-propagating lasers). As the electron moves along its trajectory, we estimate its spin polarization probability, by solving Eq.~\eqref{poability evolution}.

The solution of Eq.~\eqref{lorentz frce} is performed using different definitions of the RR force. We will refer to the `noRR'  trajectory as the solution of Eq.~\eqref{lorentz frce} with $\boldsymbol{F}_{RR}=0$, the `RR' trajectory as that  when $\boldsymbol{F}_{RR}$ is be provided by Eq.~\eqref{QRR} and the `SRR' trajectory when the equation of motion is Eq.~\eqref{SRR}.

We will also consider three different field strengths: $a_{0}=200,\;600 \;\&\;2000$, corresponding to laser intensities $I\approx5\times 10^{22},\;5\times 10^{23}\;\&\;5\times 10^{24}$ W/cm$^{2}$ for each beam, assuming 1 $\mu$m wavelength. For these intensities, the scaling laws in Ref.~\cite{del2017spin} predict the degrees of spin polarization antiparallel to $\boldsymbol{\zeta}$ of: $\lesssim$10, $\sim$50 \& $\gtrsim$70\% respectively. 

Finally we will study the robustness of the spin polarization when accounting for the instability of an electron trajectory initially displaced from the magnetic node in the z-direction.

\subsection{Influence of radiation-reaction on the electron orbits}
\label{Influence of radiation-reaction}

Let us consider an electron initially at rest at the position $k\boldsymbol{x}=(0,0,0)$, subject to the electromagnetic field in Eq.~\eqref{laser field}, with $a_{0}=600$. This laser field can be obtained from counter-propagating 1 $\mu$m wavelength laser pulses, each of intensity $I\approx5\times10^{23}$ W/cm$^{2}$.

\begin{figure*}
\subfloat[\label{traj_1.pdf}]{%
        \centering
        \includegraphics[width=0.65\columnwidth]{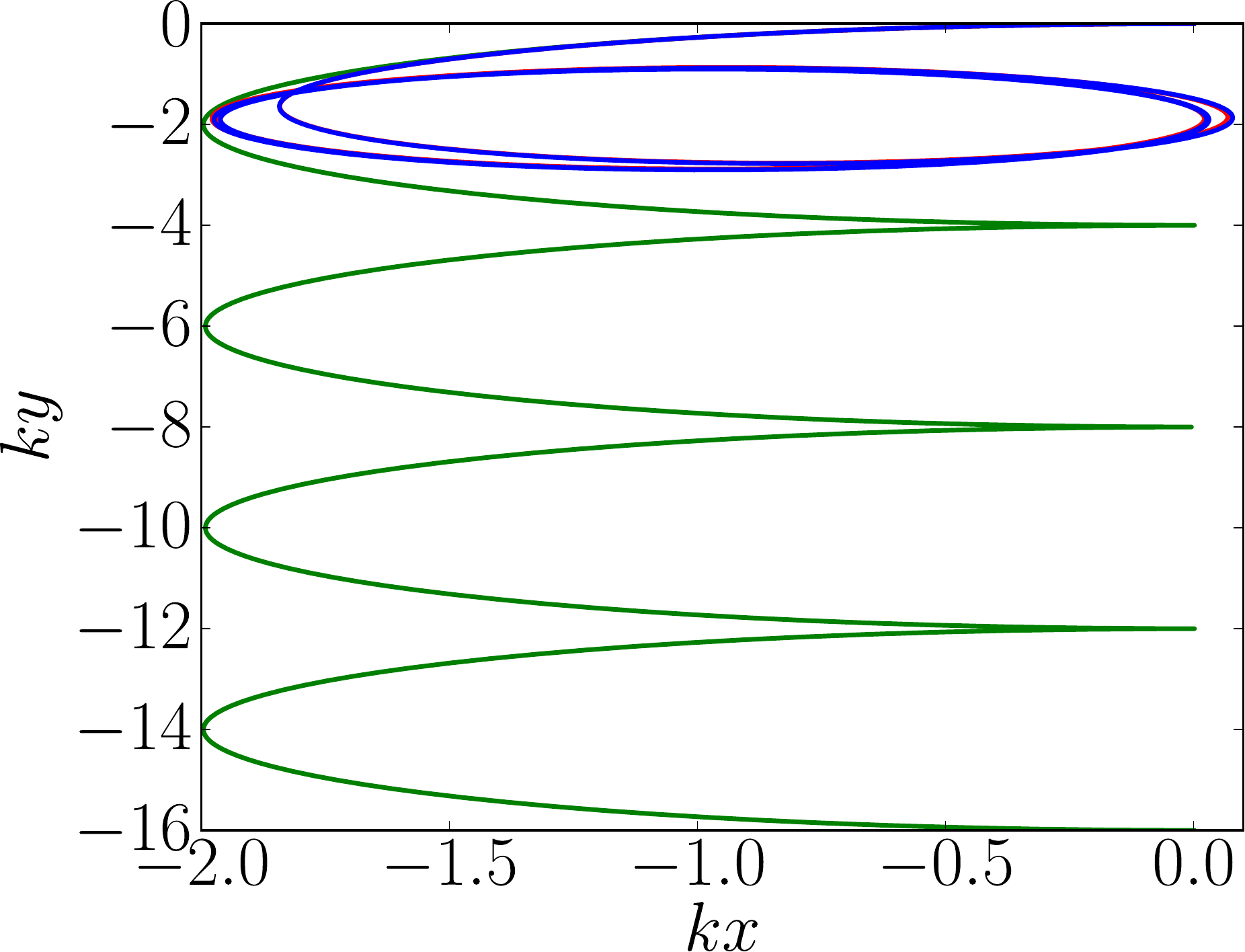}}
~
\subfloat[\label{4.pdf}]{%
        \centering
        \includegraphics[width=0.65\columnwidth]{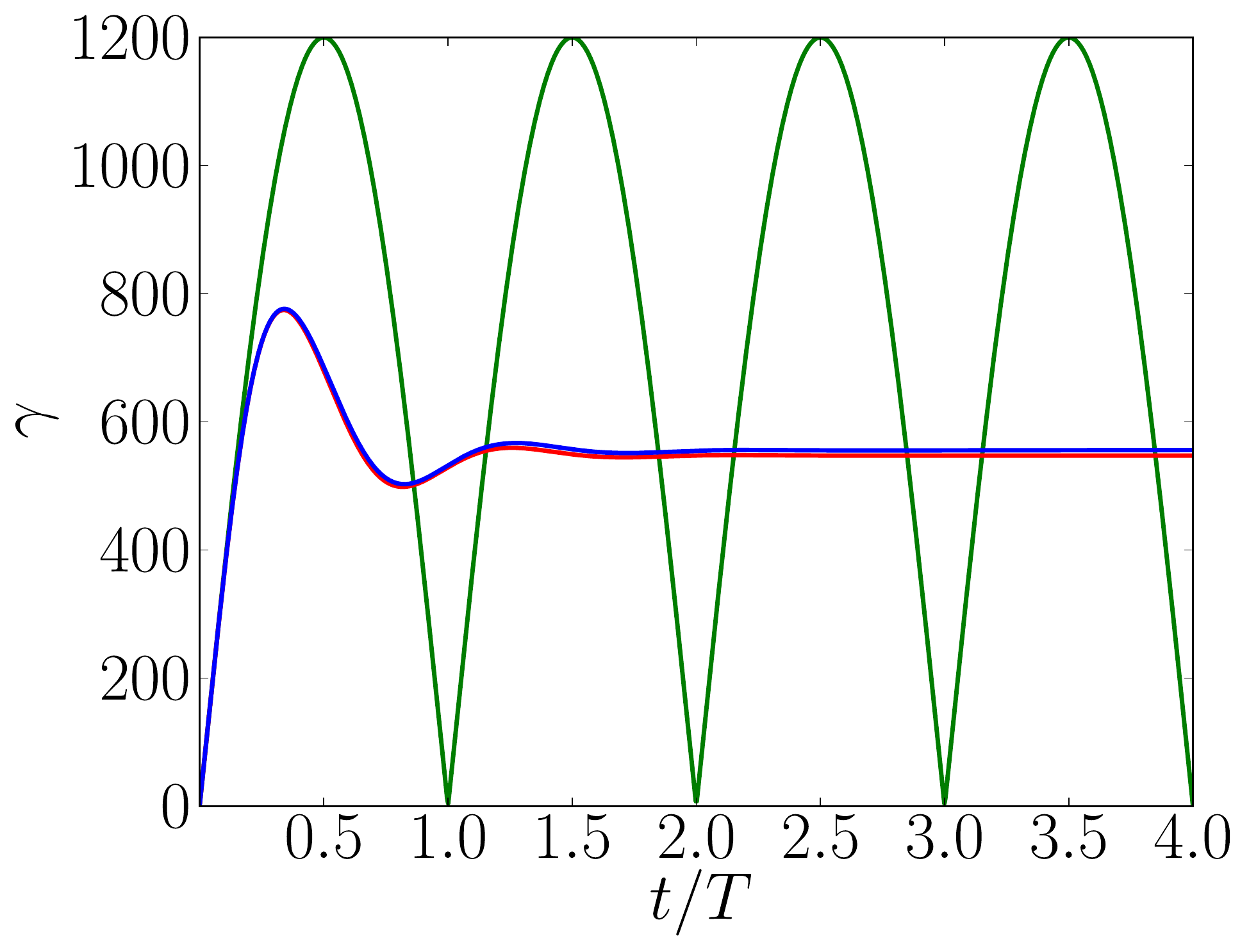}}

\subfloat[\label{5.pdf}]{%
        \centering
        \includegraphics[width=0.65\columnwidth]{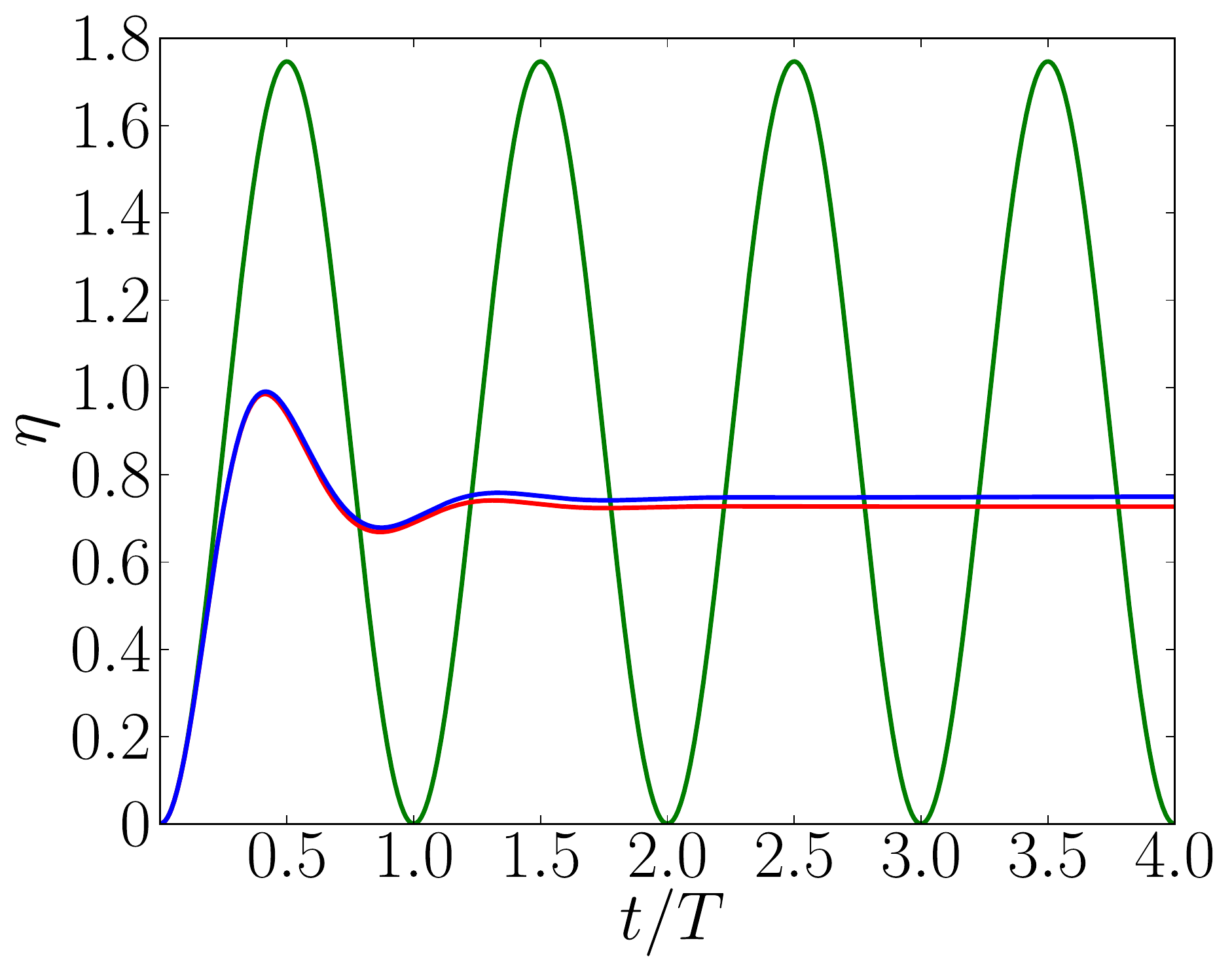}
                        }
~
\subfloat[\label{6.pdf}]{%
        \centering
        \includegraphics[width=0.65\columnwidth]{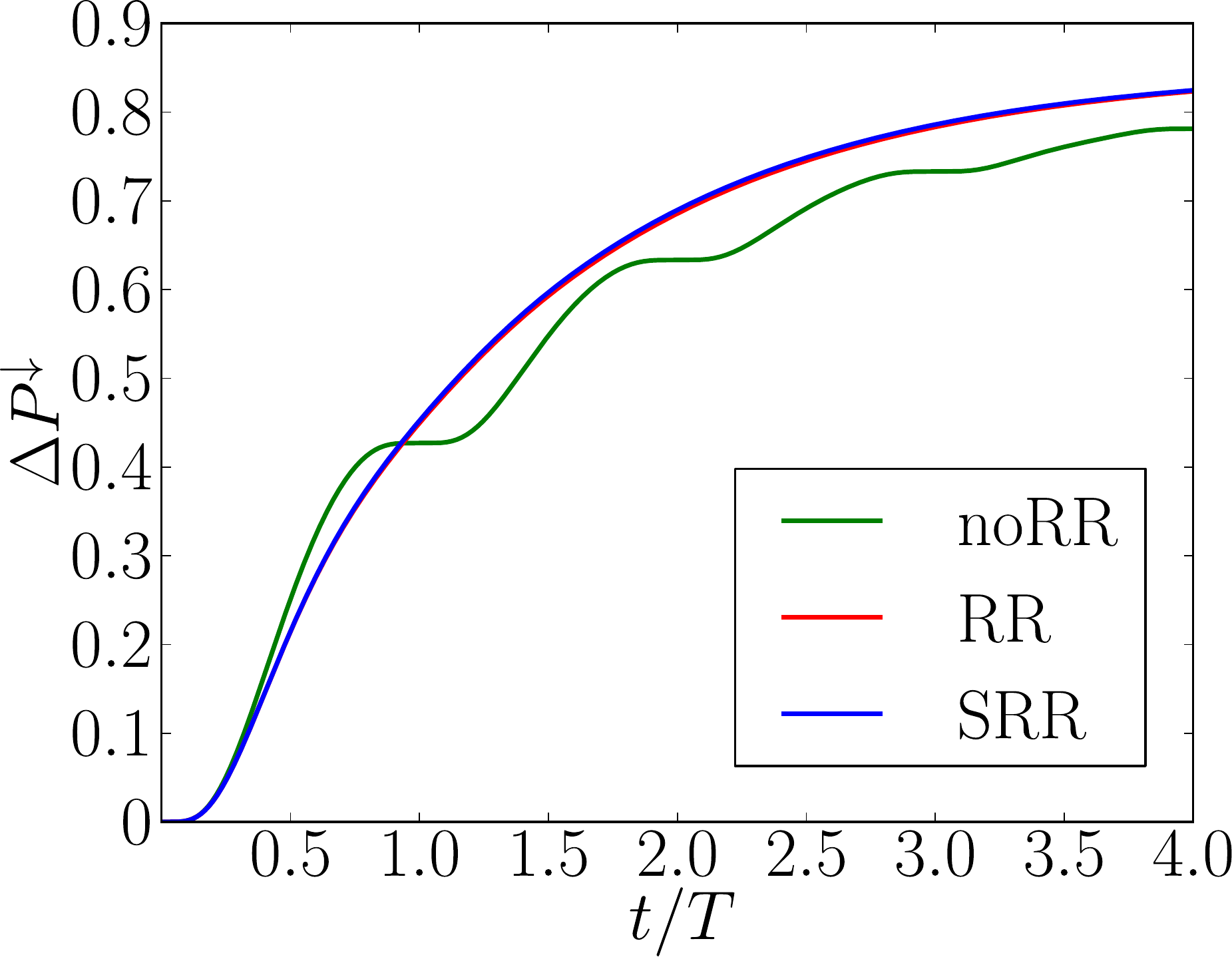}
                        }
      \caption{
      (a) Electron spatial trajectory in the magnetic node plane. (b) Its Lorentz factor, (c) non-linear quantum parameter and (d) degree of spin polarization antiparallel as functions of the time (normalized to the laser period). Three trajectories are compared: noRR (in green) RR (in red) and SRR (in blue), for $a_{0}=600$. The legend is shared among the four figures.
 }
\end{figure*}

Figure \ref{traj_1.pdf} shows the electron noRR, RR and SRR trajectories in the magnetic node plane.
As the laser accelerates the electron from rest, it very rapidly achieves ultra-relativistic velocity. Since the electric field is entirely in the $xy$ plane, no drift outside magnetic node is induced. In the noRR trajectory, the electron dynamics is characterized by a secular drift in the $y$ direction. On the contrary, when the electron energy loss due to radiation emission is considered (RR and SRR trajectories), the radiative losses due to gamma-ray emissions rapidly (in one laser period) balance the Lorentz acceleration, settling  the electron in a circular orbit. The effect of electron spin polarization on the trajectory, i.e. the difference between the RR and SRR trajectories, is small.

The noRR trajectory  is characterized by impulses at each half-period that are responsible for the secular drift. This can be seen more clearly in Figs.~\ref{4.pdf} and \ref{5.pdf}, that show $\gamma$ and $\eta$ as functions of time, for the different trajectories. At each impulse, the kinetic energy ($\propto \gamma$) reaches its maximum before decreasing. Consequently, $\eta$ oscillates between 0 and 1.8. When $\eta$ is comparable to or greater than one, spin flip induced by gamma-ray emission becomes more likely and the probability of spin polarization increases more rapidly. As is shown in  Fig.~\ref{6.pdf},  which shows the time evolution of $\Delta P^{\downarrow}$, where
\begin{equation}
 \Delta P^{\downarrow}=P^{\downarrow}-P^{\uparrow}.
\end{equation}

In the case of the RR and SRR trajectories a steady state is reached rapidly where $\gamma$ and $\eta$ stabilize to constant values with consequences on the particle degree of spin polarization antiparallel to $\boldsymbol{\zeta}$. Fig.~\ref{6.pdf} shows that in these cases $\Delta P^{\downarrow}$ increases steadily, in contrast to the noRR case that alternates rapidly increases and then plateaus due to the oscillation in $\eta$.  Despite the fact that RR and SRR trajectories are characterized by a lower $\eta$ than the maximum reached in the noRR trajectory, they reach higher spin polarization $\Delta P^{\downarrow}$ after little more than half of the laser period, due to the steady increase in the RR and SRR cases.  Note that, if we had considered a different initial condition, such as $\gamma\beta_{y}=a_{0}$, then the secular drift would also have been absent in the noRR case, suppressing the oscillation in $\eta$.  In this case we may expect the noRR case to reach higher $\Delta P^{\downarrow}$ than the RR and SRR cases.

Differences between the RR and SRR predictions of the trajectory, $\gamma$, $\eta$ and $\Delta P^{\downarrow}$ never exceed 5\% and are therefore neglected here. We conclude that the RR description of particle dynamics is sufficient and we need not consider spin effects on RR in the cases considered here. For this reason, in the rest of the article we will always solve RR trajectories only, no longer considering SRR trajectories. 

\subsection{Field strength \& initial velocity conditions}

The probability of electron spin polarization in the orbits considered here is dependent on two parameters: the time and the field strength $a_{0}$. As the electron radiates, the degree of spin polarization increases in time, reaching an asymptotic value after a time which we will call the polarization time.  As the field strength increases, both the asymptotic spin polarization and the polarization time decrease. 
In Ref.~\cite{del2017spin}, the the spin polarization time was estimated for particles initially in circular orbits. 
Here we consider an electron initially at rest.

Three characteristic laser intensities are discussed in this section: $a_{0}=200$ corresponds to the laser intensities expected to be accessible in the near term (two 1 $\mu$m wavelength lasers with $I\approx 5\times10^{22}$ W/cm$^{2}$); $a_{0}=600$, corresponding to the laser intensity which should be readily achievable with ELI \cite{korn2013eli}; $a_{0}=2000$, corresponding to $I\approx 5\times10^{24}$ W/cm$^{2}$.

\begin{figure*}
\subfloat[\label{traj_2.pdf}]{%
        \centering
        \includegraphics[width=0.65\columnwidth]{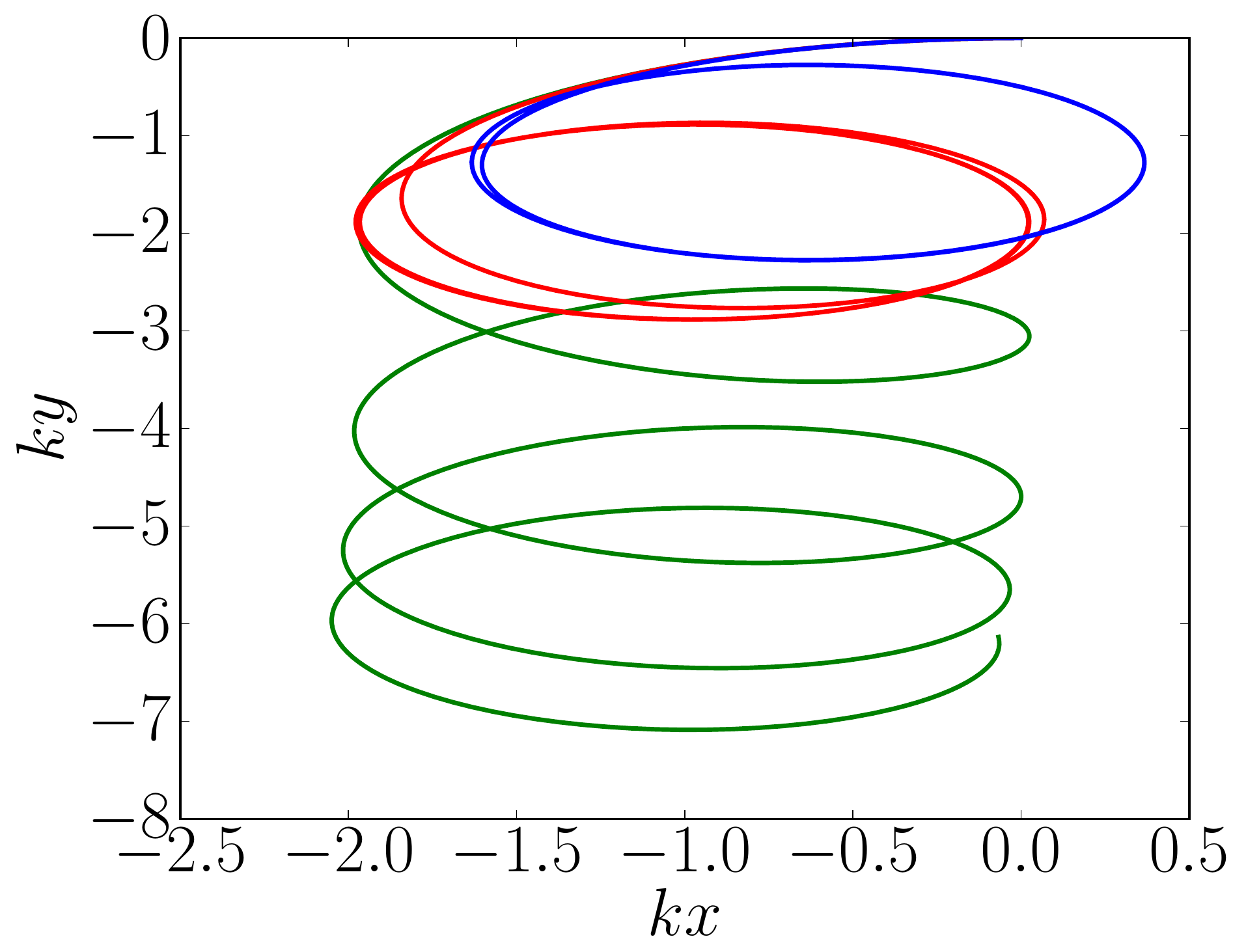}}
~
\subfloat[\label{4a.pdf}]{%
        \centering
        \includegraphics[width=0.65\columnwidth]{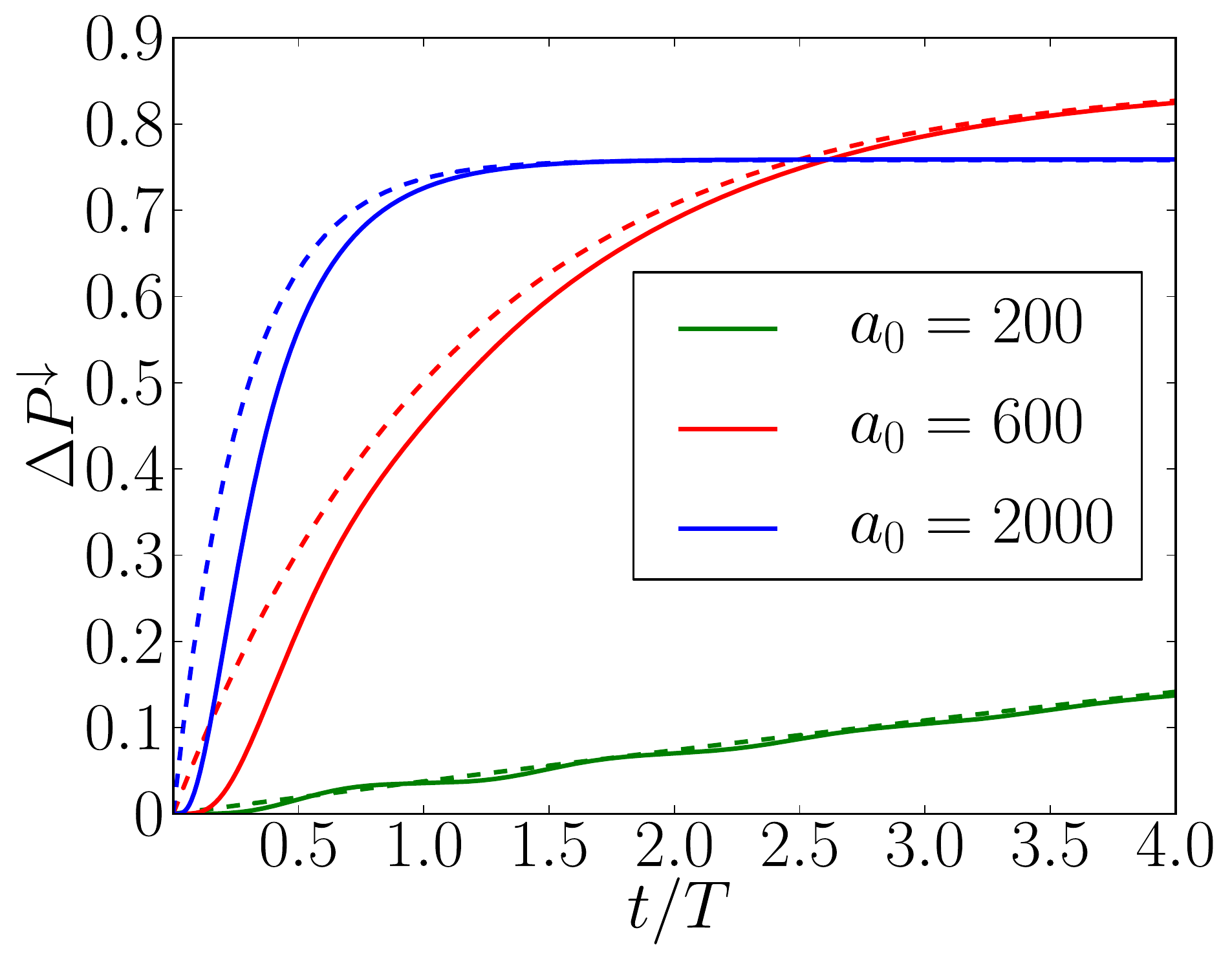}
                        }
      \caption{Spatial trajectory (a) and relative degree of spin polarization antiparallel (b) for electrons at the magnetic node of two counter-propagating laser fields with $a_{0}=200,$ 600 \& 2000. Continuous lines refer to electrons initially at rest and dashed lines to electrons settled in the circular trajectory from the outset.  The legend is shared among the figures and the simulation time is $4T$.
 }
\end{figure*}

The RR spatial trajectory of electrons initially at rest and subject to the field strengths $a_{0}=200$, 600 and 2000 are shown in Fig.~\ref{traj_2.pdf}. The trajectory for $a_{0}=600$ has already been discussed in Sec.~\ref{Influence of radiation-reaction}. 
For $a_{0}=200$, the electron dynamics is characterized by weak radiation emission. For this reason the electron does not settle into the circular orbit in the time considered, after which it is still drifting in the $y$ direction (although the drift velocity is decreasing in time).
On the contrary, for $a_{0}=2000$, the RR force is very strong and sufficient to settle the electron in the circular orbit in less than half a period. 

In Ref.~\cite{del2017spin} predictions of the spin polarization were made for the three cases described here, with the electrons settled in the circular orbit from the outset. In figure \ref{4a.pdf} we compare the degree of spin polarization for these predictions to the case where the electron starts at rest, i.e. the three trajectories considered in figure \ref{traj_2.pdf}. For $a_{0}=200$, the degree of spin polarization antiparallel to $\boldsymbol{\zeta}$ is small ($\Delta P^{\downarrow}\leq 20\%$) because $\eta$ is small ($\eta<0.18$). Therefore the probability of emission and therefore spin flip is reduced. In the case of electrons initially in the circular orbit, $\Delta P^{\downarrow}$ increases linearly with time as the time-scale considered is much shorter than the time for the spin polarization to saturate -- the polarization time.  If the electron is initially at rest, the growth of $\Delta P^{\downarrow}$ is characterized by the alternation of relatively rapid growth and then a plateau due to the periodicity in $\eta$ caused by the drift in the $y$-direction, as shown in Sec.~\ref{Influence of radiation-reaction}. Nevertheless the difference in $\Delta P^{\downarrow}$ between the cases where the electron is initially in the circular orbit and where it starts from rest is small. 

For  $a_{0}=600$, $\Delta P^{\downarrow}$ increases linearly for approximately $0.5T$ and then begins to saturate as the polarization time (which is shorter for $a_0=600$ than for $a_0=200$) is approached. For this laser intensity, the difference in $\Delta P^{\downarrow}$ between electrons initially at rest and those initially in the circular orbit is small and decreases in time. The difference is due to the fact the the electron initially settled in the circular orbit is initially more energetic that the electron initially at rest, so it has a higher probability of radiating and therefore of spin flip.  For $a_{0}=2000$, after one laser period, $\Delta P^{\downarrow}$ reaches its asymptotic value.  The electron initially at rest very quickly reaches the circular orbit and therefore, differences between electrons initially at rest or already in their circular trajectory are even smaller  than in the case with $a_{0}=600$.

In this section  we have shown that the spin polarization of electrons at the magnetic node of the laser configuration discussed in this article is weakly dependent on their initial velocity.  Electrons initially at rest will obtain a smaller degree of spin polarization than those initially in a circular trajectory but the difference never exceeds 10\% and decreases rapidly (over one laser period) in time.

\subsection{Trajectory instability \&  spin precession}

At present, we do not have a good model to describe the spin flips (radiative polarization) and the classical spin precession simultaneously.
Therefore, the theory for electron spin polarization discussed in this article and in Ref.~\cite{del2017spin} relies on the choice of a globally non-precessing spin basis. 
Such a basis is represented by the vector parallel to the magnetic field in the electron's instantaneous rest frame. For the laser configuration discussed, a globally non-precessing spin basis can only be found if the electron's motion is confined to the magnetic node: the vector $\boldsymbol{\zeta}=\boldsymbol{e}_{z}$.

\begin{figure}{
        \centering
        \includegraphics[width=1.\columnwidth]{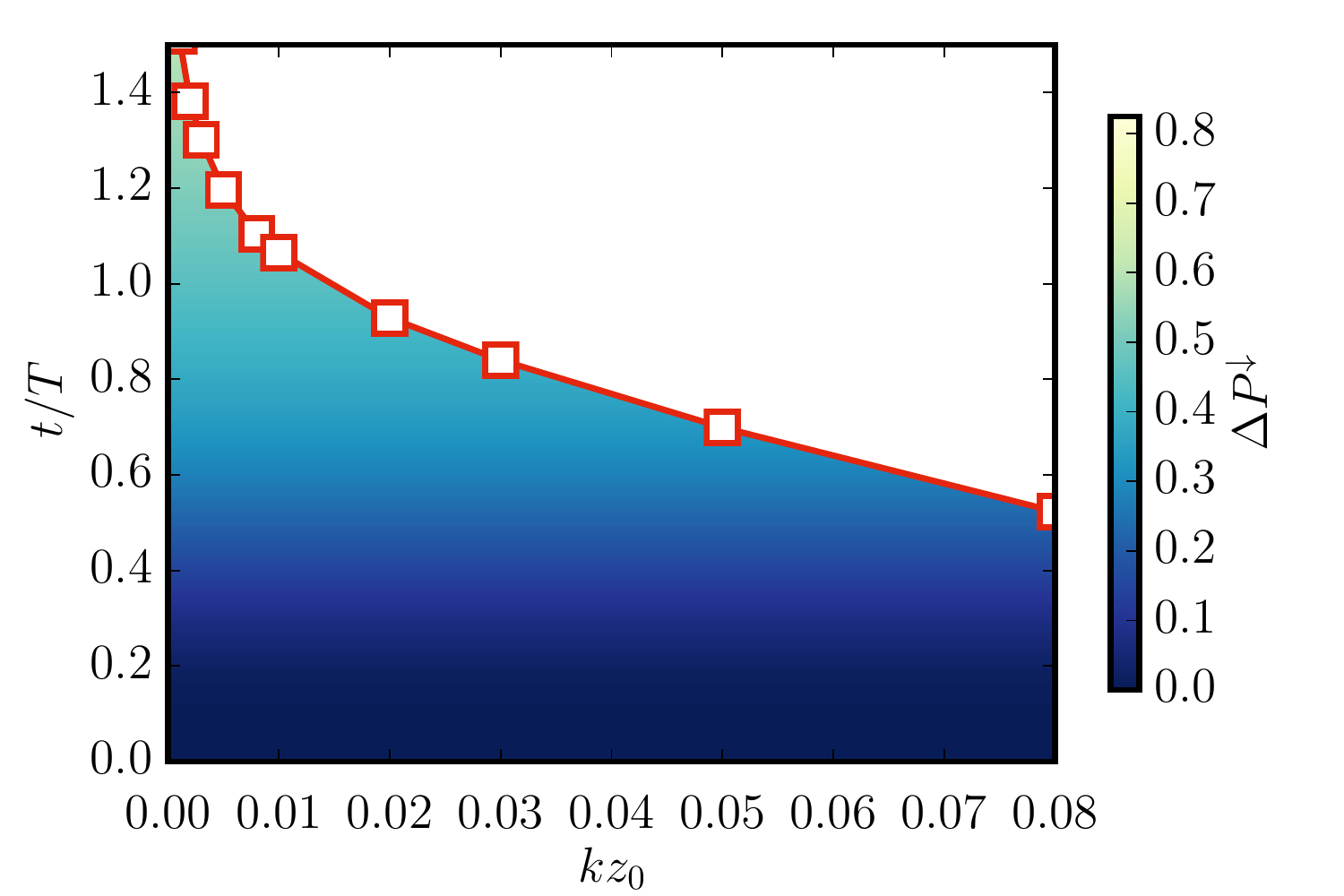}
          }    
      \caption{Spin precession time as a function of the electron initial position (off the magnetic node $z=0$). As color-plot, also the degree of spin polarization as a function of time is shown. Each point represents a different simulation. \label{daniel_600_contour.pdf}
 }
\end{figure}

In Ref.~\cite{kirk2016radiative}, the stability of the trajectory at the magnetic node has been discussed in detail. It has been shown that it is an unstable trajectory: any small perturbation from that position giving an irreversible deviation from the position of the node. This deviation happens in a timescale of the same order as the laser period. We now aim to determine the consequences, for the degree of spin polarization, of a small deviation from the magnetic node. We consider the case of an electron initially at rest, in the standing wave created by lasers with $a_{0}=600$ at a position a few percent of a laser wavelength away from the magnetic node.

As the electron migrates away from the magnetic node plane, $\vec \zeta=\vec e_{z}$ is no longer a non-precessing spin basis. Classically, the spin expectation value starts to precess, according to Eq.~\eqref{BMT eq}.
In Fig.~\ref{daniel_600_contour.pdf}, we plot the time for the spin expectation value to undergo significant precesssion -- the precession time. Note that we do not include the additional radiative spin polarization resulting from spin flip transitions, only the classical spin precession, due to the lack of an appropriate model. The spin is assumed initially parallel to $\vec e_{z}$ and the precession time is defined as the time the expectation value of the spin takes to differ by 50\% from its initial value.
This value is computed considering the spin in the particle instantaneous rest frame, related to its laboratory counterpart by the transformation
\begin{equation}
 \boldsymbol{S}_{RF}=\boldsymbol{S}-\frac{\gamma}{\gamma+1}\boldsymbol{S}\cdot\boldsymbol{\beta}\boldsymbol{\beta}.
 \end{equation}
  The precession time is plotted as a function of the initial position $kz_{0}$ (off the magnetic node $kz=0$). 
\color{black}

The spin precession time represents an upper limit for the validity of our theory, i.e. we consider the predictions of the simple model presented here to be valid only on a timescale shorter than this. This gives an upper limit on the spin polarization for electrons off the magnetic node which is shown by the color map in Fig.~\ref{daniel_600_contour.pdf}.  This shows that we can expect a degree of spin polarization higher than 30\%, for  $a_0=600$, for electrons initially $\lesssim 1\%$ of a laser wavelength off the magnetic node. Analogously, for $a_{0}=2000$, we expect a 70\% degree of spin polarization for electrons initially $\lesssim 1\%$ of a laser wavelength off the magnetic node.

The electron migration from the magnetic node will quickly suppress  spin flip transitions, due to the decrease in $\eta$ caused by the decrease in the electric field away from the node. This suggests that away from the magnetic node the spin polarization achieved before migration may be preserved. On the contrary, the electron's trajectory is chaotic as it migrates and therefore the spin expectation value would be expected to precess chaotically, causing a depolarization.

\section{Discussion}\label{Discussion}

Recent progress in the study of RR in experiments with high intensity lasers \cite{cole2017experimental,poder2017evidence,wistisen2017experimental} motivates the need to recognize signatures of quantum effects on RR. The spin polarization of electron beams may be one of the clearest, because it has no classical counterpart. 
However, further work is needed to accurately model the spin polarization dynamics for realistic experimental conditions

 The model we used in this article to describe electron spin polarization is based on a deterministic description of the particle trajectory and on a stochastic description of spin flip transitions along a non-precessing polarization direction. Using this approach we have examined the robustness of electron spin polarization induced by ultra-intense lasers to effects such as the instability of the magnetic node and variations in the initial velocity of the electron.

We have seen that the electron dynamics is weakly affected ($\lesssim 5\%$) by the polarized nature of the particle. On the contrary the RR force plays an important role
because it settles electrons in a stationary orbit, such that $\eta$ is constant and $\sim1$. In this case, electrons radiate continuously in time, decreasing the polarization time.
 The RR force also helps to confine the electron in the magnetic node. In particular, this second statement could be important when considering lasers with a realistic focal spot size.

One major limitation of this analysis is our inability to deal with spin precession and spin flip simultaneously as the electron migrates away from the magnetic node (an effect of far greater importance that variations in the initial velocity of the electron -- which causes the spin polarization to vary by $\lesssim$5\%).  However, classical tracking of the spin precession indicates that
 the spin polarization time is shorter than the time electrons take to migrate from the magnetic node: we expect a systematic polarization of the magnetic node plane region, due to the quick polarization of incoming electrons. 
 The spin polarization should be preserved for electrons within $0.01$ laser wavelengths of the magnetic node suggesting that thin targets may be advantageous. \color{black}  Multi-dimensional effects such as laser focusing will further complicate this picture and have not been considered here.  We may be able to find additional configurations where spin precession is suppressed by considering more complex field configurations where the electrons may be trapped in a rotating field \cite{kirk2016radiative}.

In this article we have used a deterministic model for the radiation reaction force (as described in \cite{ridgers2017signatures}).  In the quantum regime the emission is a stochastic process, however it has recently been shown that a semi-classical model reproduces the ensemble average behavior of an electron population well \cite{ridgers2017signatures,niel2017quantum} and thus we would expect it to predict the spin polarization (which is an expectation value) well.  Stochasticity may affect the rate of migration of the electrons from the magnetic node, an effect which warrants further investigation. 

\color{black}

To include electron and positron spin dynamics in simulations of next generation laser-plasma interactions correctly a model which can describe spin flip in arbitrary fields is required.  Developing such a model is important as the electron spin polarization should modify the polarization of the radiated gamma-ray photons, which could modify the dynamics of electron-positron cascades \cite{king2013photon} and these can play a crucial role in next generation laser matter interactions \cite{del2018efficient,grismayer2017seeded}.

\section{Conclusions}\label{Conclusions}

In this article we have discussed the possibility of electron spin polarization in realistic trajectories around the magnetic node of the standing wave set up by two circularly polarized, counter-propagating, ultra-intense lasers. We have characterized the conditions in which we can confidently expect an important degree of electron spin polarization. A significant degree ($>$5\%) of spin polarization can be expected for $a_{0}\gtrsim200$ and that the instability of the electron trajectories at the magnetic node gives an upper limit to the achievable spin polarization as the spin precesses as the electrons migrate from this unstable point.  The possibility of producing spin-polarized electrons with ultra-intense lasers paves the way for new applications. Polarized electrons are fundamental for the study of particle physics and are used in the spin polarized electron spectroscopy. 
\color{black}

The data required to reproduce the results in this article are available from the University of York at DOI 10.15124/b25e6428-ae40-43fb-b91d-f2785a09b5bc.

\section*{Acknowledgments}
This work was funded by the UK Engineering and Physical Sciences Research Council (Grant No. EP/M018156/1) and by
the Science and Technology Facilities Council (Grant No. ST/G008248/1).  The authors are grateful to J. G. Kirk, S. Meuren and A. Di Piazza for useful and stimulating discussions.

%

 \end{document}